\documentclass[a4paper,11pt]{article}

\usepackage{jinstpub} 
\usepackage{graphicx}
\usepackage{lineno}
\usepackage{amsmath}
\DeclareGraphicsExtensions{.jpg,.pdf,.png}

\newcommand{\cevns}{CE$\nu$NS}
\newcommand{\gf}{\textsc{Geant4}}

\newcommand{\Co}{$^{60}$Co}
\newcommand{\Cs}{$^{137}$Cs}
\newcommand{\Tl}{$^{208}$Tl}

\title{Characterization of the ambient background in the RED-100 experiment location at Kalinin Nuclear Power Plant}

\abstract{The RED-100 experiment with a liquid xenon target was carried out at Kalinin Nuclear Power Plant. The goal of the experiment is the detection and study of the coherent elastic neutrino nucleus scattering process (\cevns{}) for the low-energy antineutrinos in close vicinity to a reactor core. A good understanding of the external radioactive background is needed to achieve this goal. This paper describes the external background conditions for the RED-100 experiment at Kalinin Nuclear Power Plant.}


\begin{document}
\author[a]{D.Y.~Akimov,}
\author[a,b]{I.S.~Alexandrov}
\author[a,c]{V.A.~Belov,}
\author[a]{A.I.~Bolozdynya,}
\author[a,c]{A.V.~Etenko,}
\author[a,d]{A.V.~Galavanov,}
\author[a,d]{Yu.V.~Gusakov,}
\author[a,b]{A.V.~Khromov,}
\author[a,e]{A.M.~Konovalov,}
\author[a,f]{V.N.~Kornoukhov,}
\author[a,c]{A.G.~Kovalenko,}
\author[a]{E.S.~Kozlova,}
\author[a,b]{A.V.~Kumpan,}
\author[a]{B.O.~Lavrov,}
\author[a,g]{A.V.~Lukyashin,}
\author[a]{A.V.~Pinchuk,}
\author[a,c]{O.E.~Razuvaeva,}
\author[a,1,\dag]{D.G.~Rudik,\note{Corresponding author} \note[\dag]{Now at: University of Naples Federico II, Corso Umberto I 40, Naples, 80138, Italy}}
\author[a]{A.V.~Shakirov,}
\author[a,c]{G.E.~Simakov,}
\author[a]{V.V.~Sosnovtsev,}
\author[a]{A.A.~Vasin}

\affiliation[a]{National Research Nuclear University ``MEPhI'' (Moscow Engineering Physics Institute)\\ 31 Kashirskoe hwy, Moscow 115409, Russia}
\affiliation[b]{National Research Tomsk Polytechnic University\\
30 Lenin ave, Tomsk, 634050, Russia}
\affiliation[c]{National Research Center “Kurchatov Institute”\\
1 Akademika Kurchatova sq., Moscow, 123182, Russia}
\affiliation[d]{Joint Institute for Nuclear Research\\ 6 Joliot-Curie St, Dubna, Moscow region 141980, Russia}
\affiliation[e]{P.N. Lebedev Physical Institute of the Russian Academy of Sciences\\ 53 Leninskiy Prospekt, Moscow, 119991, Russia}
\affiliation[f]{Institute for Nuclear Research\\7a 60-letiya Oktyabrya ave, Moscow, 117312, Russia}
\affiliation[g]{Russian Technological University, Lomonosov Institute of Fine Chemical Technologies\\ 86 Vernadsky Avenue, Moscow, 119571, Russia}

\emailAdd{rudik.dmitry@mail.ru}
\maketitle
\flushbottom

\section{Introduction}
\label{sec:intro}

The process of coherent elastic neutrino nucleus scattering (\cevns{}) was predicted more than 45 years ago~\cite{kopeliovich1974isotopic,freedman1974coherent}, but was observed only recently by the COHERENT experiment~\cite{akimov2017observation,akimov2020argon}. According to the Standard Model of elementary particles (SM), the cross section of this process depends quadratically on the number of neutrons in the nuclei. Therefore, for the heavy nuclei, the \cevns{} cross section is by two orders of magnitude higher than the cross section of the inverse beta decay. The prevailing cross section of \cevns{} over all other known neutrino interactions makes this process very interesting as a possible tool for nuclear reactor monitoring and nonproliferation tasks~\cite{bernstein2020colloquium}. On the other hand, the small energy deposition of \cevns{} is challenging to detect~\cite{akimov2019coherent}.

There are several experiments around the World which are trying to measure \cevns{} at reactors~\cite{bonet2021constraints, colaresi2021first, agnolet2017background, angloher2019exploring, salagnac2023optimization, flores2021physics, wong2018taiwan, belov2015nugen, aguilar2020search}. RED-100 has the largest sensitive mass among other \cevns{} experiments at reactors, and it is the only detector with liquid xenon as a target. As with all other detectors, RED-100 meets extreme conditions at a reactor site. For example, very high temperature variations with reactor operation can cause the instability of electronics threshold levels and, consequently, result in deviations in background rate. 

For the reactor \cevns{} experiments it is very important to measure the ambient background. Reactor correlated background can cause events in the detectors which can mimic \cevns{} events~\cite{hakenmuller2019neutron}. It also should be noted that one could not just use the same background spectra and rates obtained during the reactor OFF period in order to get the background estimation during reactor ON since the rates and spectra could be different. Thus, independent continuous monitoring of different components of background is important.

In this paper, we describe the result of ambient background measurements and monitoring during the RED-100 data taking period at the Kalinin Nuclear Power Plant (KNPP). In section~\ref{sec:red100}, a short description of the RED-100 experiment and the experimental site is given. Section~\ref{sec:gamma} is devoted to the gamma background measurements and monitoring. Section~\ref{sec:neutron} is about neutron background. In section~\ref{sec:radon}, we discuss the limits on possible radon background. In section~\ref{sec:muon}, we describe the main background in the region of interest (ROI) for RED-100 caused mainly by muons. We also show in this section our measurement of primary muons flux and the monitoring of the main background count rate. Finally, conclusions are presented in section~\ref{sec:conclusion}.

\section{The RED-100 experiment at KNPP}
\label{sec:red100}

RED-100 is a two-phase liquid xenon detector that was built to detect \cevns{} in close vicinity of reactor core~\cite{akimov2020first}. It was deployed under the 4th block of Kalinin Nuclear Power Plant (KNPP) at 19 m from the center of the active zone. The detailed description of the RED-100 setup can be found, for example, in ref.~\cite{akimov2022red}. In this section, a brief overview of the experimental setup is performed with a focus on ambient background monitoring.

RED-100 is located at the ground level, two levels below the active core of the standard 3GW thermal power WWER-1000 reactor unit. The estimated shielding from the cosmic background in a vertical direction is about 50 meters of water equivalent (m.w.e.)~\cite{alekseev2016danss}. Also, this location is well shielded from the reactor itself by the biological shield, the thick ceilings of two levels, and the moderately high distance from the reactor.

The passive shield of RED-100 consists of 5 cm of copper and about 70 cm of water in all directions. The study of passive shielding efficiency and its detailed description can be found elsewhere~\cite{akimov2021passive}. Although this shielding suppress the ambient gamma background at least two orders of magnitude according to our previous study and it is almost opaque to the low energy neutrons, it is important, nevertheless, to monitor the fluxes of external backgrounds to monitor possible difference in count rate of the detector between the periods of Reactor ON and OFF.

Four background monitoring detectors were installed and continuously operated during the RED-100 data taking period: two domestic radon indicators, the NaI[Tl] detector for gamma background monitoring, and the Bicron liquid scintillator (BC501A) detector for fast neutron background monitoring. The latter two detectors were placed close to the water tank of the RED-100 passive shield at $\sim$180 cm above the floor, approximately at the level of the RED-100 sensitive volume center.

Also, several additional campaigns were provided before and during the RED-100 operation to measure and characterize the gamma and radon background in place. The gamma background was characterized with the bigger NaI[Tl] detector (see next section), which was used in our previous laboratory measurements~\cite{akimov2021passive}. The radon background was measured by the KNPP staff several times during the RED-100 operation. According to these measurements, the radon background was below the sensitivity of their detectors which is 20 Bq/m$^3$.

\section{Gamma background}
\label{sec:gamma}

There were two independent sets of ambient gamma background measurements. The first one was provided before RED-100 and the supporting structure was deployed. The scintillator detector with NaI[Tl] cylindrical crystal of height 10 cm and diameter 15 cm was used to scan the experimental hall to find possible gamma background hot spots around the future detector location. Its characterization and performance were described in detail in~\cite{akimov2021passive}. During RED-100 data taking period, the gamma background was continuously monitored with a smaller detector with NaI[Tl] crystal of 8 cm height and 8 cm diameter. It was located constantly at a height of about 180 cm from the floor, attached to the supporting frame of the RED-100 detector.

In both cases, an independent electronics rack was used for the power supply and data taking. Signals from NaI[Tl] were amplified and shaped with the ORTEC 572A NIM unit with further digitization with ORTEC 927 MCA. The laptop with original MAESTRO Software was used to record and store amplitude spectra acquired during 20-minute long runs. These spectra were analyzed then offline.

\begin{figure}[htbp]
    \centering
    \includegraphics[width=.4\textwidth]{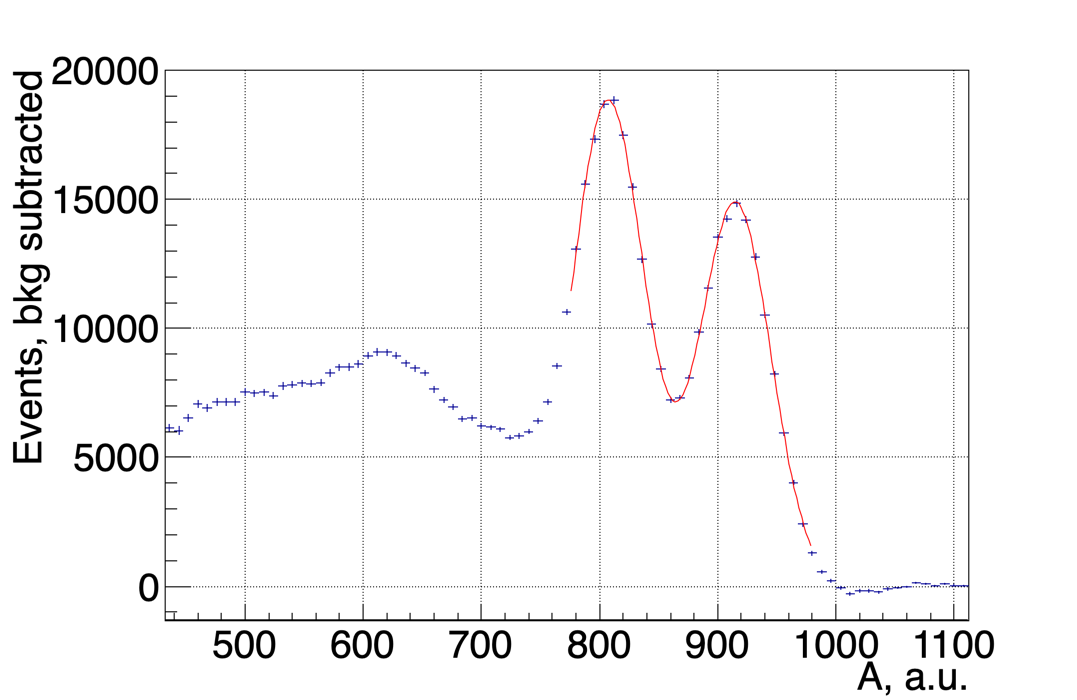}
    \qquad
    \includegraphics[width=.4\textwidth]{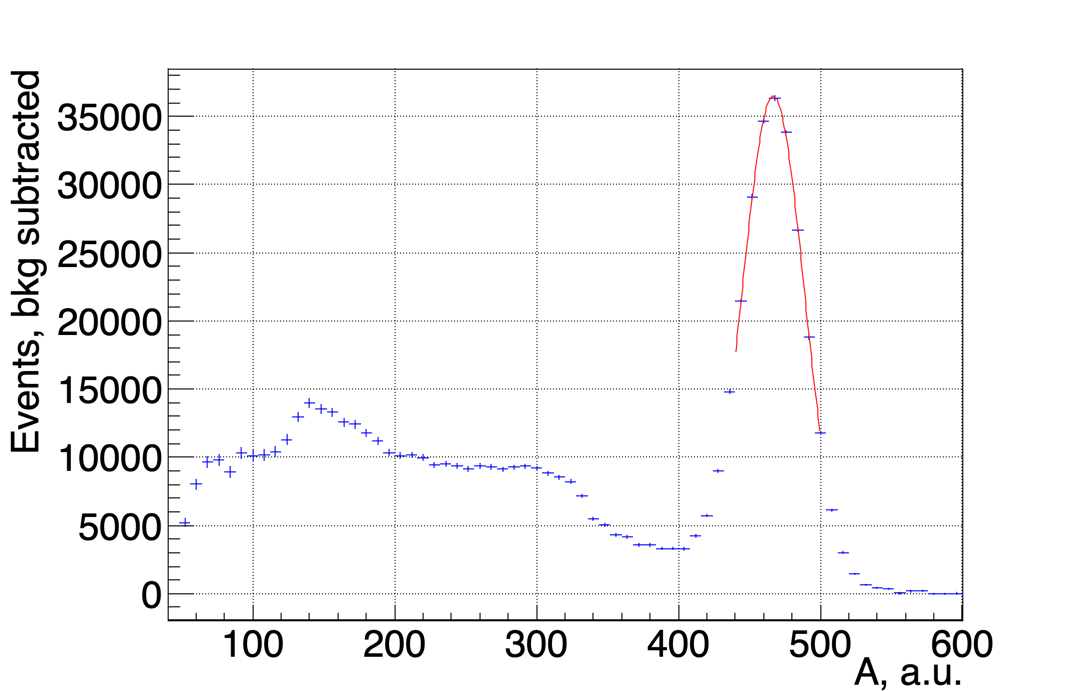}
    \caption{\label{fig:nai_sources}NaI[Tl] spectra from gamma sources: \Co{} on the left and \Cs{} on the right; background is subtracted.}
\end{figure}

To check the stability of the response and to obtain the detector energy scale, NaI[Tl] was calibrated weekly with \Co{} and \Cs{} sources.
Examples of spectra from these sources with subtracted backgrounds are given in figure~\ref{fig:nai_sources}. The light yield stability based on weekly calibration monitoring of the detector was at a level of $\sim 2\%$ through all data taking period. This number was taken into account for the systematic uncertainty estimation for the count rate.

\begin{figure}[htbp]
    \centering
    \includegraphics[width=.4\textwidth]{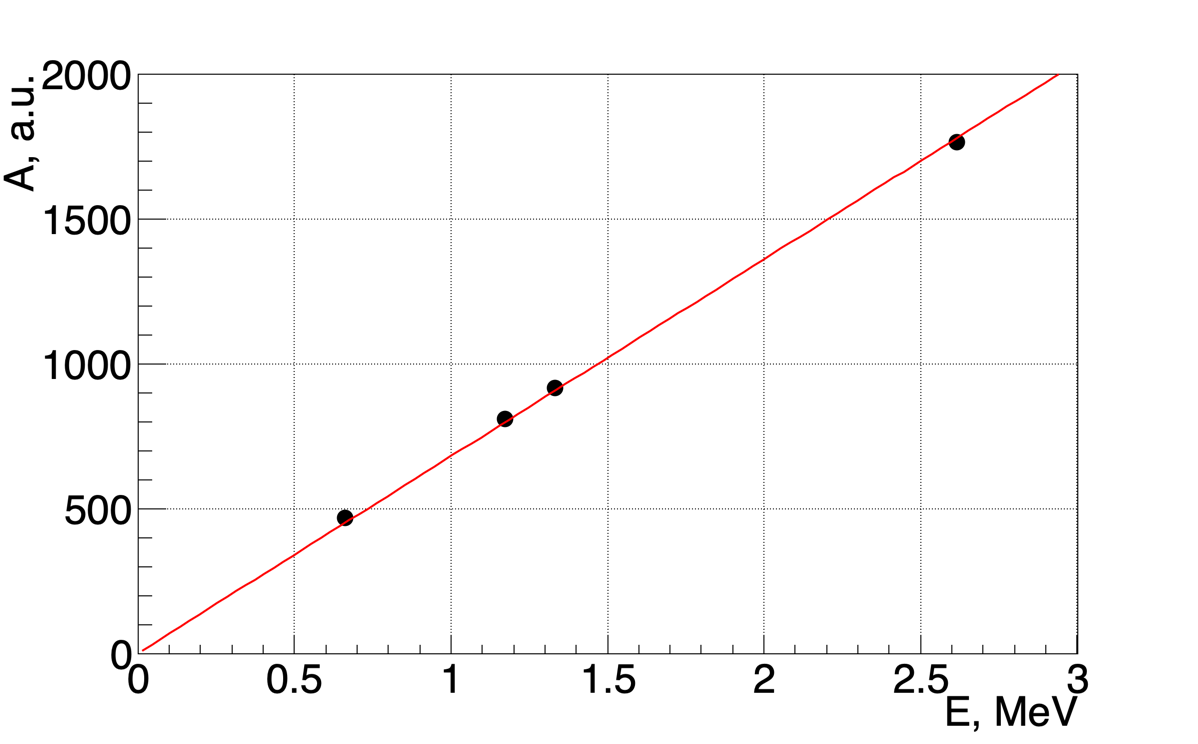}
    \qquad
    \includegraphics[width=.4\textwidth]{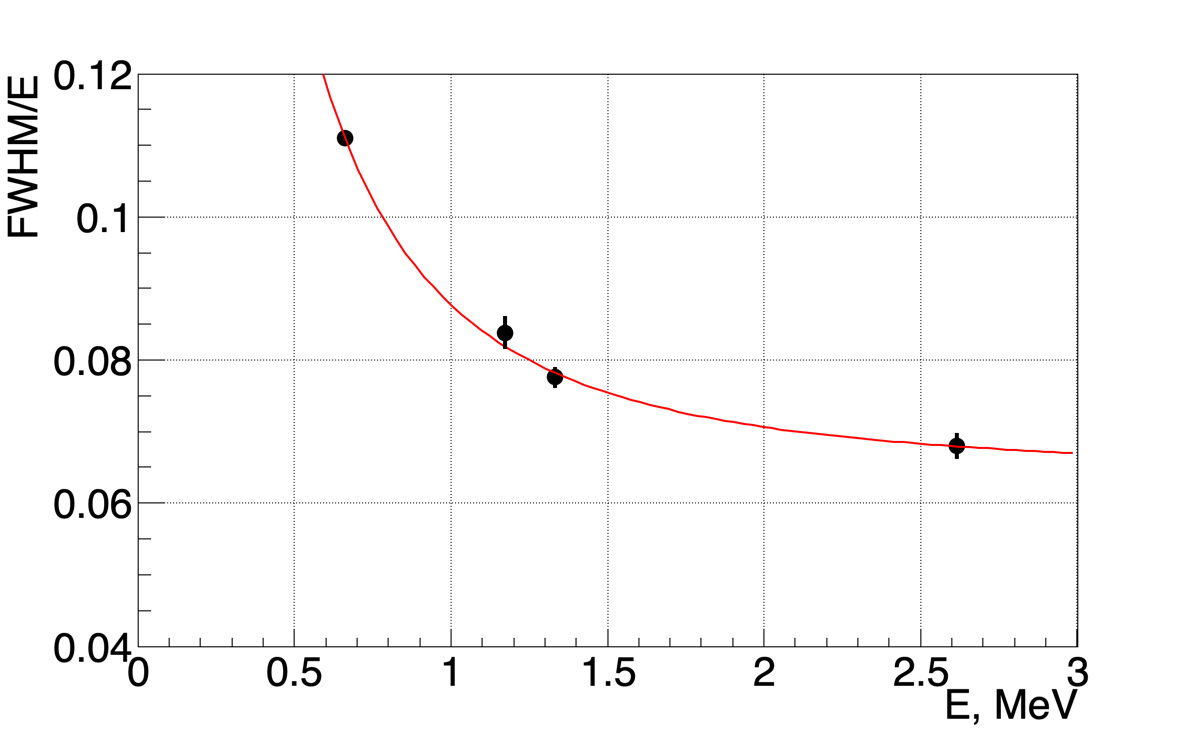}
    \caption{\label{fig:nai_calib_res}NaI[Tl] calibration and resolution}
\end{figure}

To extend the energy calibration range, the line of \Tl{} in the natural background with maximum energy was used.
The example of NaI[Tl] calibration and energy resolution plots is presented in figure~\ref{fig:nai_calib_res}. Good linearity of the detector response in the energy range from 0.5 to 2.6 MeV was obtained. The detector resolution at the line of \Cs{} is approximately 11\%, which is enough for background monitoring purposes.

The energy resolution points were fitted according to the formula~\ref{eq:nai_resolution}:
\begin{equation}
    \centering
    \frac{FWHM}{E} = \sqrt{a^2 + \left(\frac{b}{\sqrt{E}}\right)^2 + \left(\frac{c}{E}\right)^2},
    \label{eq:nai_resolution}
\end{equation}
where $a$, $b$ and $c$ represent constant, stochastic, and noise terms with the obtained values $5.8\%$, $3.5\%$, and $5.5\%$, respectively. This energy resolution dependence was incorporated into the \gf{}~\cite{agostinelli2003geant4} Monte Carlo (MC) model in order to determine components of the background spectrum.

\begin{figure}[htbp]
    \centering
    \includegraphics[width=.4\textwidth]{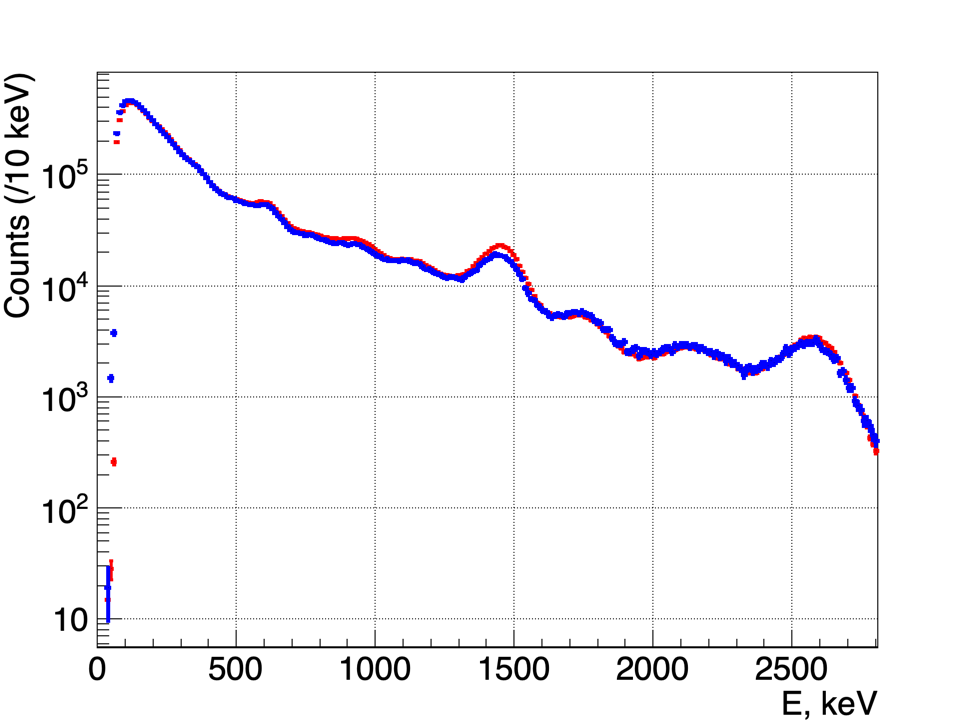}
    \qquad
    \includegraphics[width=.415\textwidth]{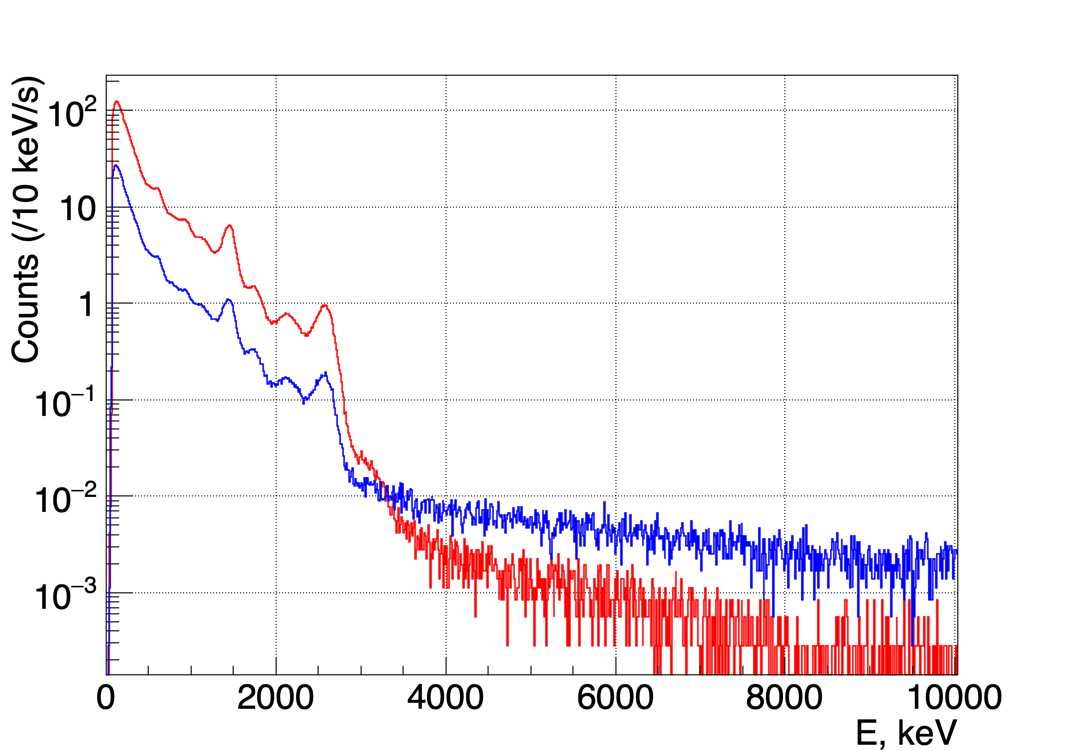}
    \caption{Comparison of the gamma NaI[Tl] spectrum obtained at KNPP (in red) with the one obtained in the laboratory measurements (in blue); on the left plot, normalization is performed by total integral, on the right plot, the spectra are normalized by count rate.}
    \label{fig:nai_knpp_vs_lab}
\end{figure}

The first set of measurements in different locations of the experimental hall has shown that the gamma background is mostly natural. In figure~\ref{fig:nai_knpp_vs_lab}, there is a comparison between spectra obtained at KNPP (in red color) and during the laboratory tests at MEPhI (in blue color). Deviations in the $^{40}$K peak height can be explained by slightly different content of this isotope in the concrete at the laboratory and KNPP. The count rate at KNPP is by a factor of 4.8 higher than that in the laboratory tests due to the much thicker concrete floor, ceiling, and walls. A lower count rate in the high energy region in the KNPP data is associated with the lower muons rate at KNPP due to almost 50 m.w.e. provided by the power unit building and the reactor itself above the experimental hall.

\begin{figure}[htbp]
    \centering
    \includegraphics[width=.8\textwidth]{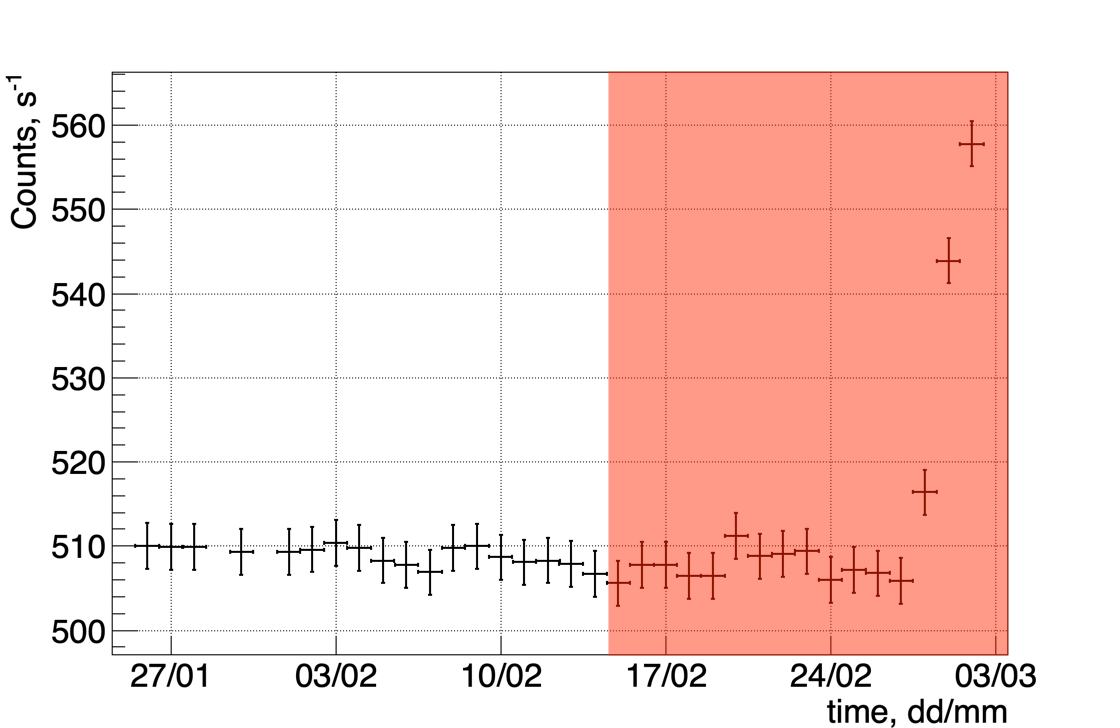}
    \caption{Averaged count rate per day during RED-100 experimental run; red transparent rectangle marks the reactor ON period; the increased count rate at the end of the monitoring cycle corresponds to the water draining from RED-100 passive shielding.}
    \label{fig:nai_count_rate}
\end{figure}

During the RED-100 experimental run, the small NaI[Tl] detector was used for continuous external background monitoring. The spectra, acquired over 20 minutes each, were stored during the detector operation. Calibration of the detector with gamma sources described above was done once a week for one hour per source. To get the plot of count rate versus time, calibration periods and bad-quality 20-minute data collecting bunches were excluded. These bad-quality data files originated rarely by manually switching off the data acquisition loop before calibrations or by the routine restart of the loop and their rate did not exceed one bad 20-minute file per 2--3 days. Then, the total amount of collected events was normalized to the data collecting live time for each day. The result of the average count rate per second per whole energy range (0.05 -- 10 MeV) during each day is in figure~\ref{fig:nai_count_rate}.

As can be seen from figure \ref{fig:nai_count_rate}, there was no influence of the reactor operation on the gamma background in the RED-100 experimental hall. The increasing count rate trend that started at the end of February was caused by the draining of the water from the RED-100 passive shield.

Knowing that the count rate is independent of the reactor operation status, having spectrum and rate in hand, the RED-100 response on the gamma background can be simulated. The proportion of background components (K$^{40}$, U$^{238}$ and Th$^{232}$) is considered to be the same as in our previous measurements~\cite{akimov2021passive}. Simulation of the RED-100 response for the background components will be published in upcoming papers.

\section{Neutron background}
\label{sec:neutron}

One of the most important backgrounds in reactor neutrino experiments is the neutron background. Neutrons can mimic \cevns{} signal in the ROI. It can provide a reactor-correlated background since the neutron flux may increase with the reactor turning on as it was shown in previous experiments~\cite{hakenmuller2019neutron}.

The RED-100 experimental site is well shielded against slow neutrons
originating in the reactor by the biological shield of thick concrete and by additional thick concrete ceilings at the level above. Background measurements at the level above can be found in paper~\cite{alekseev2016danss}. The RED-100 passive water shield of 70 cm water in all directions eliminates neutrons background with energies below 1 MeV.

Fast neutrons can be produced by cosmic muons interacting with the building constructions around the detector. As soon as different manipulations with water levels in the reactor and fuel water pools occur during the reactor OFF period, one might expect variations in the muon flux which can provide variations in the neutron flux as well.

To measure and monitor the fast neutron background, a liquid scintillator BC501A Bicron was used. An important feature of a liquid scintillator detector is the presence of at least two components in the scintillation light with significantly different decay times. The ratio between fast and slow components intensity depends on the interaction particle type~\cite{vcerny2004study}. Thus, it is possible to use pulse shape discrimination (PSD) techniques to distinguish neutrons and gamma interactions~\cite{guerrero2008analysis}.

\begin{figure}[htbp]
    \centering
    \includegraphics[width = .4\textwidth]{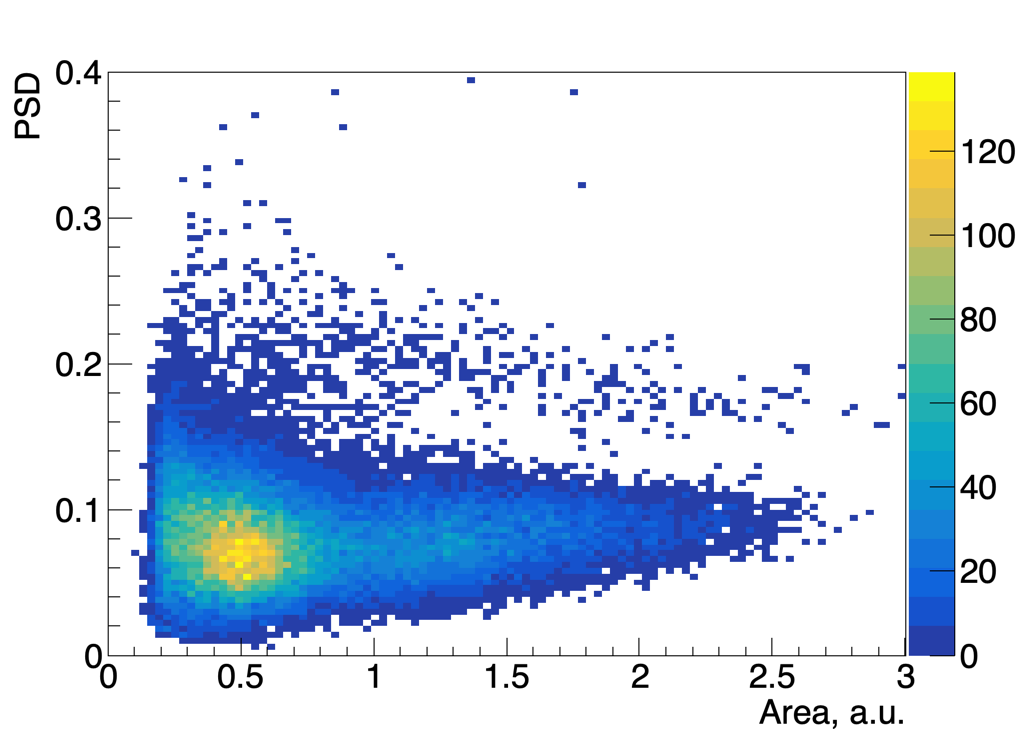}
    \qquad
    \includegraphics[width = .4\textwidth]{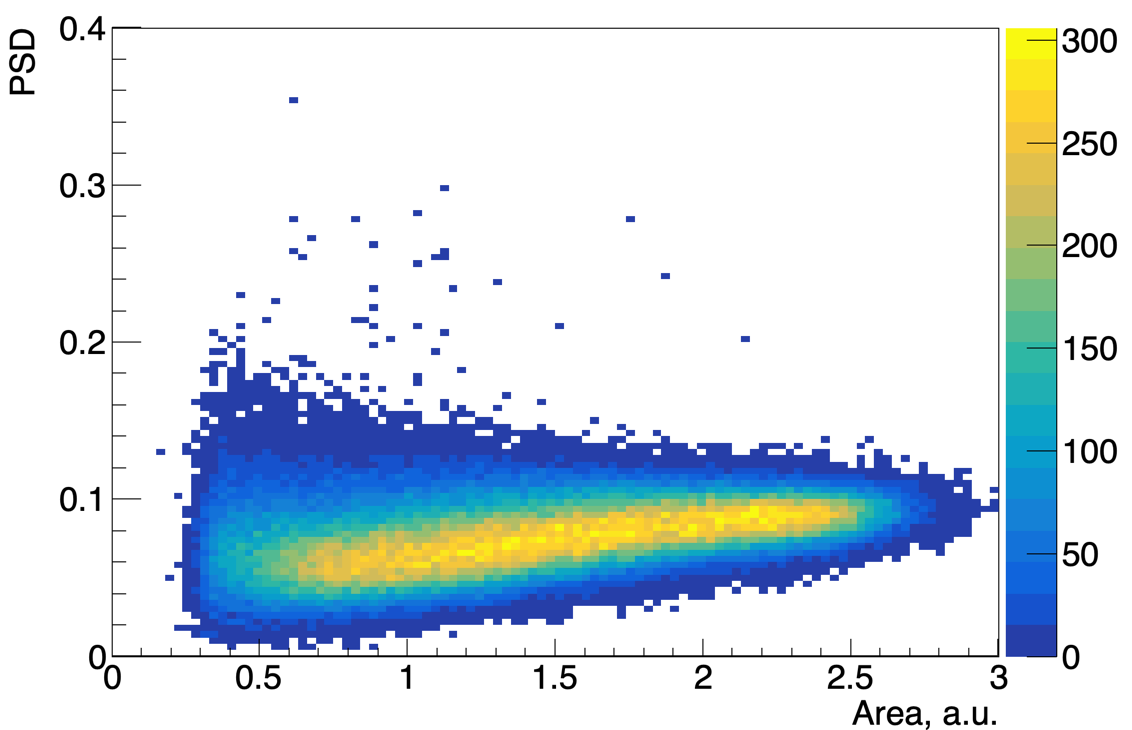}
    \caption{Dependence of PSD parameter on the area of the scintillation signal for the PuBe source (left) and the background at the RED-100 location (right).}
    \label{fig:bicron_psd_example}
\end{figure}

To check the ability of our detector to distinguish neutrons from gammas, specially dedicated measurements were done in an adjoining room with reactor monitoring equipment containing a PuBe source. Two scatter plots in figure~\ref{fig:bicron_psd_example} represent measurements with and without PuBe source. PSD parameter was defined as the area of scintillation signal obtained at the event tail to the total event area. To increase the discrimination power of the PSD method the threshold between the event tail and the signal onset was varied taking the resolution between two bands as a figure of merit. The best threshold was found at 23 ns from the beginning of the scintillation signal.

In figure \ref{fig:bicron_psd_example}, two separate bands are clearly seen for the data obtained with and without the PuBe source. Only a few neutron events were observed for the same period of time without a PuBe source which can be associated with neutrons generated by cosmic muons at the detector location. Different energy spectra of gamma background at different sites explain the difference in intensities of gamma bands.

\begin{figure}[htbp]
    \centering
    \includegraphics[width = .4\textwidth]{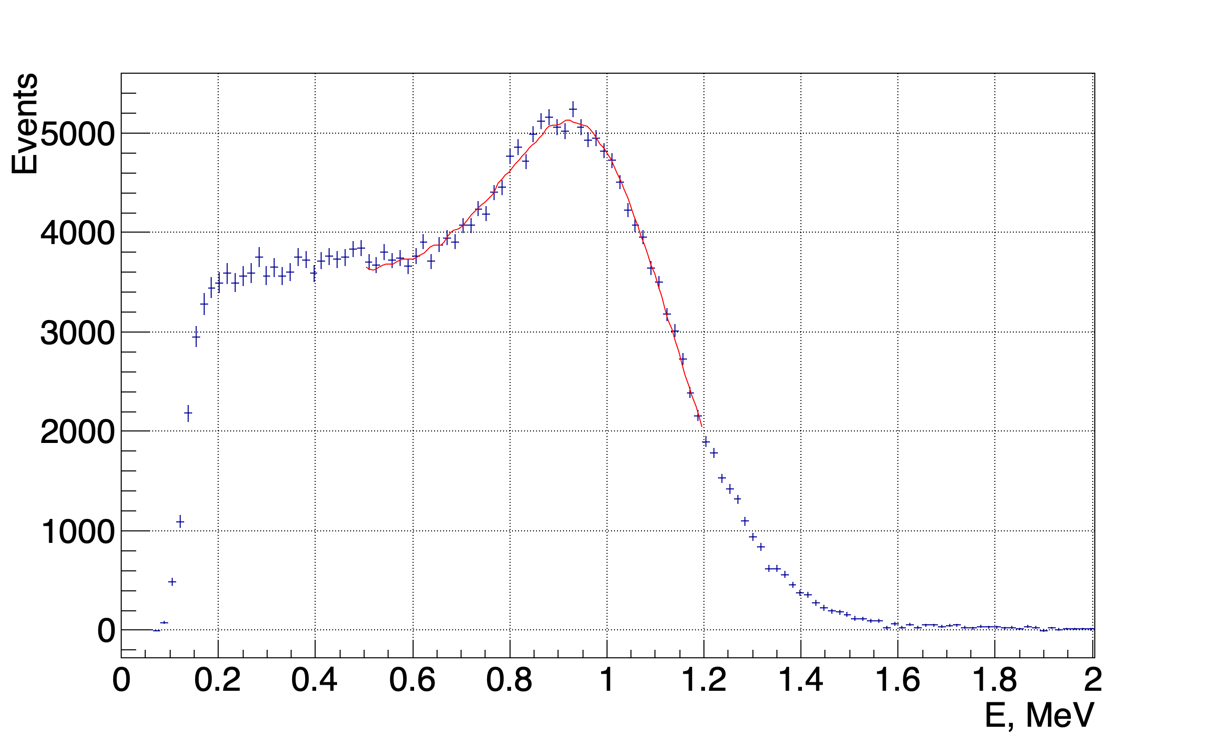}
    \qquad
    \includegraphics[width = .4\textwidth]{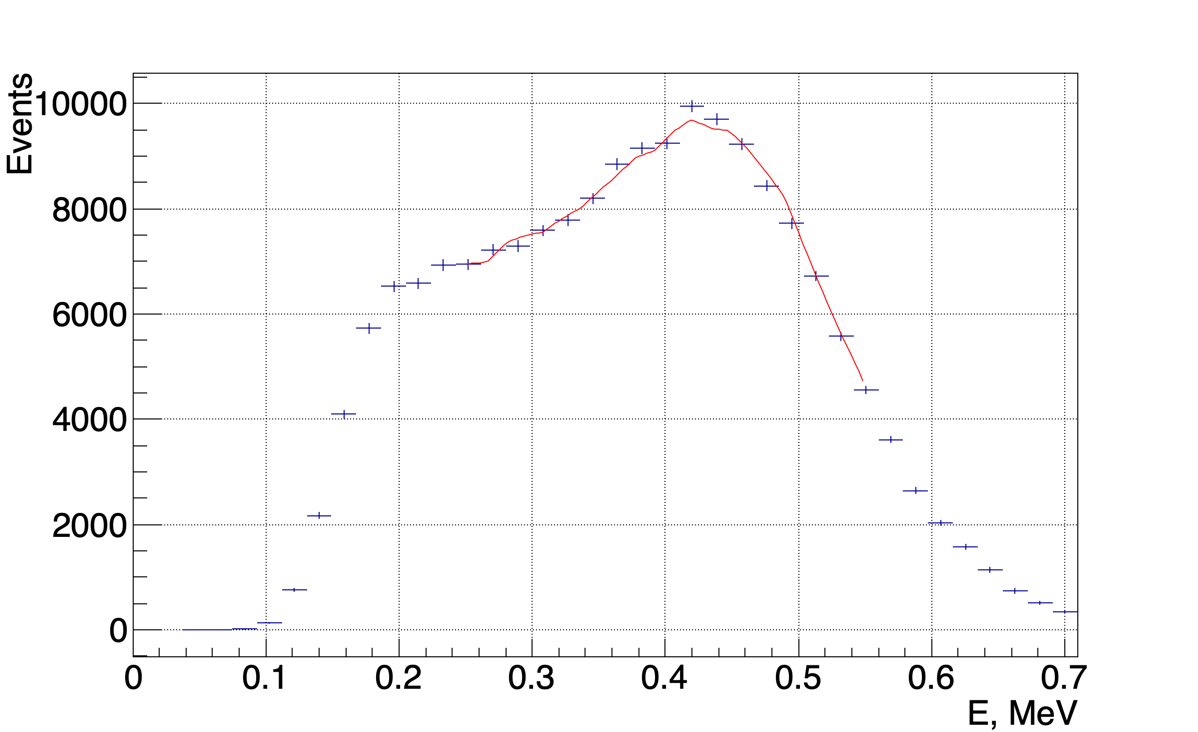}
    \caption{Calibration data obtained with \Co{} (left) and \Cs{} (right) sources with the \gf{} produced approximation.}
    \label{fig:bicron_calibration}
\end{figure}

Since the BC501A liquid scintillator is a light material it is hard to calibrate it using gamma sources~\cite{klein2002neutron}. 
There are no photo peaks in the pulse height spectra obtained with gamma sources, only Compton edges are visible. Therefore, to perform a fit of the \gf{} simulated spectra to the experimental ones, the energy scale and resolution for each calibration source were varied simultaneously and iterative to minimize the $\chi^2/ndf$ fit parameter. The final versions of MC fit to the data for two gamma sources \Co{} and \Cs{} are shown in figure~\ref{fig:bicron_calibration}.

A similar procedure was done to fit the gamma background spectrum in the energy region of the \Tl{} Compton edge. The spectra of the main background components (K$^{40}$, U$^{238}$ and Th$^{232}$ radioactive chains) were simulated. The background spectrum was fitted with all 3 components simultaneously using TFractionFitter ROOT class~\cite{barlow1993fitting, brun1997root}. The $\chi^2/ndf$ parameter was minimized in the iteration process to get the best fit of the measured background spectra with the simulated one.

\begin{figure}[htbp]
    \centering
    \includegraphics[width = .4\textwidth]{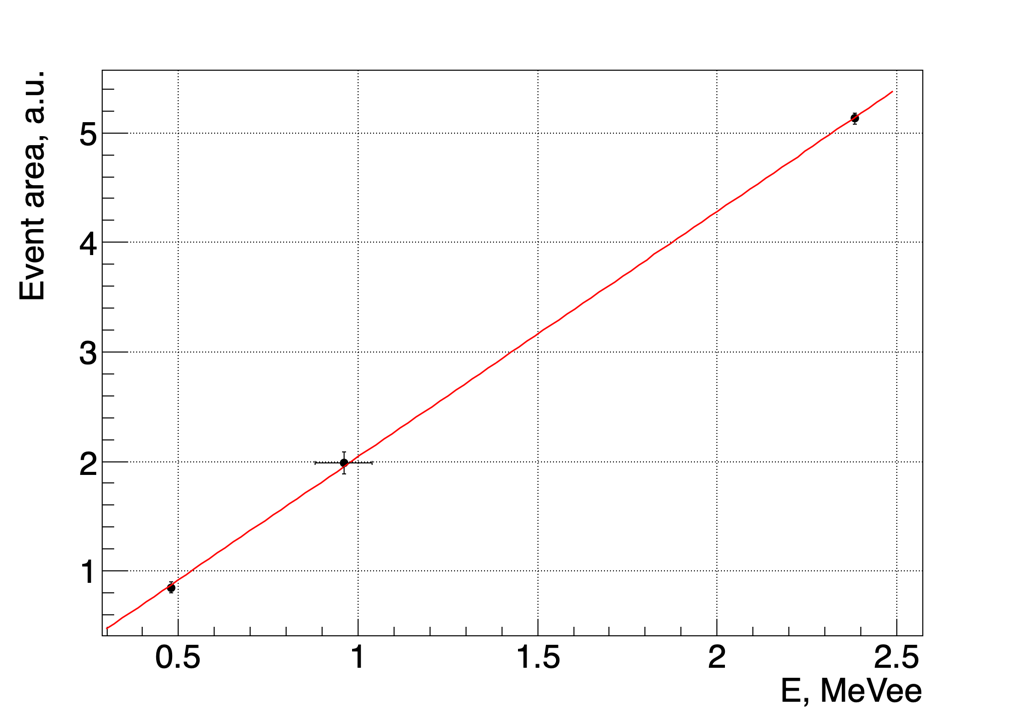}
    \qquad
    \includegraphics[width = .4\textwidth]{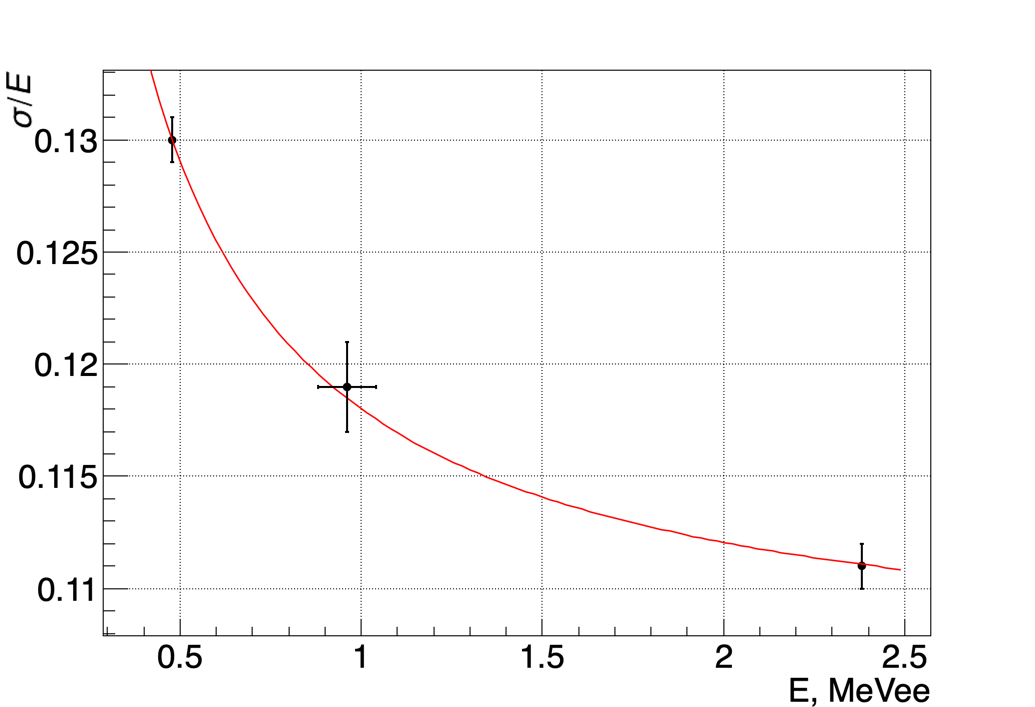}
    \caption{Bicron scintillation detector calibration and resolution obtained with gamma sources and background data; the energy uncertainty for the \Co{} calibration point is enhanced to fulfill the fact of the presence of two unresolved Compton edges for this source.}
    \label{fig:bicron_calib_resolution}
\end{figure}

Finally, a correspondence between the original event area obtained in the analysis and energy in MeV of electron recoils (MeVee) was established (see figure ~\ref{fig:bicron_calib_resolution} on left) and the dependence of energy resolution ($\sigma$) was obtained (see figure ~\ref{fig:bicron_calib_resolution} on right). Good linearity of the detector response was achieved for the whole calibration range. The stability of the Bicron detector during the data taking period was checked with a weekly calibration procedure. It was observed that the value of light yield decreased slightly during data taking reaching a maximum deviation of $\sim5\%$. This was accounted for the systematic uncertainty estimation.

To predict the neutron background in RED-100 one should know the neutron flux and the energy spectrum at the detector site. Since the energy deposition for neutrons and electrons of the same energy differs in liquid scintillators, one should use the quenching factor (QF) for re-scale energy from MeVee to MeV nuclear recoils (MeVnr). We use QF parametrization from ref.~\cite{arneodo1998calibration}. This parametrization along with the measured detector resolution was incorporated into \gf{} model of the Bicron detector. With this model, a procedure of energy unfolding and flux normalization was performed.

The limited experimental statistics of neutron events don't allow us to characterize possible features on the neutron energy spectrum. Thus, an unfolding procedure was simplified to get only a general behavior of the energy spectrum as a power law function of neutron energy~\cite{sanna1971monte, reginatto2010overview, matzke2003unfolding}. 

\begin{figure}[htbp]
    \centering
    \includegraphics[width = 0.8\textwidth]{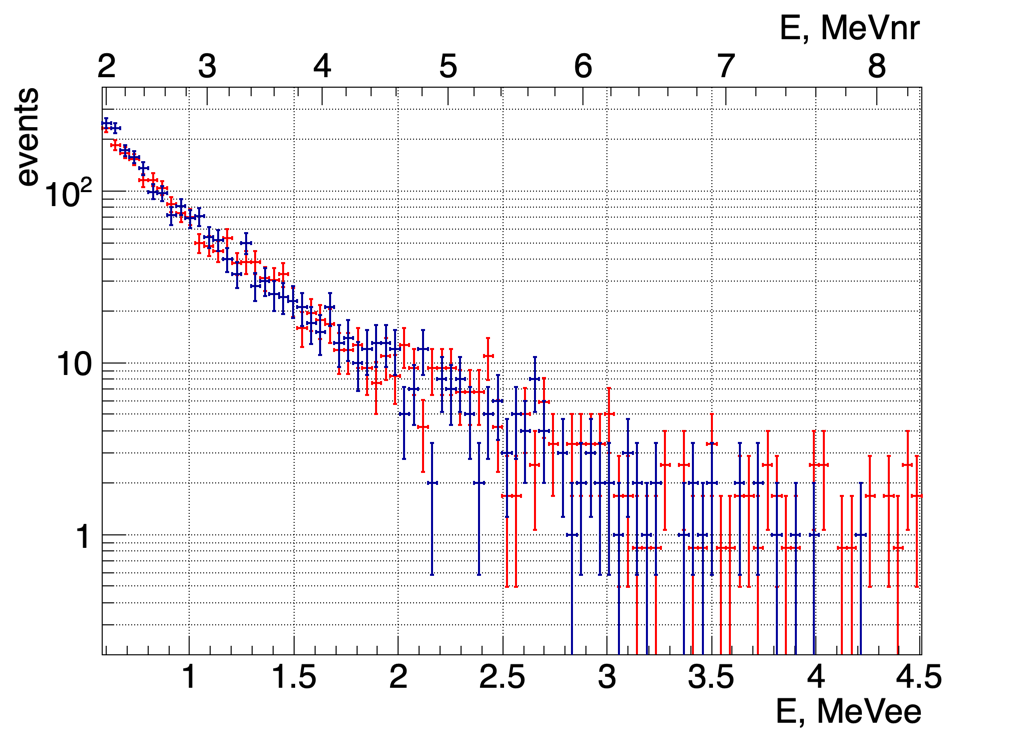}
    \caption{Experimental neutron energy deposition spectrum (blue) and the MC simulated one corresponded to a power-law neutron spectrum with $E^{-2.9}$; axis at the top represents energy in MeVnr scale with QF taken into account.}
    \label{fig:bicron_spectrum}
\end{figure}

To find the power law which provides the best fit of the simulated spectrum to the data the following iterative algorithm was used. Neutrons with a power-law energy spectrum were randomly generated in the \gf{} model from the flat square source of $2*2$ meters, located at a distance of one meter from the edge of the detector. The angular distribution of the neutrons was isotropic in a solid angle of $2\pi$ to the same side where the detector was placed. The energy deposition for the neutron events was quenched according to the QF parametrization and additionally smeared by the obtained detector resolution. Then, the data was fitted with the simulated spectra in the energy region 0.6 -- 4 MeVee. A low energy cut was used to avoid the region of low PSD power in order to avoid the reduction of efficiency due to the  PSD cut. The best fit of the power-law spectrum was determined by minimizing $\chi^2/ndf$ and is equal to $- 2.9 \pm 0.1$. The experimental neutrons spectrum was collected through all the data taking periods and the final MC spectrum is presented in figure~\ref{fig:bicron_spectrum}.

The number of MC neutrons, passed through the flat square counter of 2*2 cm, located at the central axis above the detector model in parallel with the source plane, was determined. Having this number of neutrons in hand, and the scaling factor from fitting the MC spectrum to the experimental one, the neutron flux at the experimental site normalized to the obtained neutron spectrum can be calculated. The problem is that in the energy range 0 -- 0.6 MeVee the neutrons and the electrons bands on the PSD plot are overlapping and it is impossible to calculate the exact number of neutrons in this region. At the same time it can't be lower than the number of neutrons passed PSD cuts, so this number was set as a lower limit of the neutron flux and was $(9.2\pm0.5) \times 10^{-5} neutrons/cm^2/s$. An upper limit was set in an assumption that the neutron spectrum continues to follow the $E^{-2.9}$ power law down to the neutron energy of about 1 MeVnr ($\sim0.2$ MeVee), which is at the edge of the Bicron sensitivity. Thus, an upper limit of neutrons flux was estimated as $(24.1\pm 1.2) \times 10^{-5} neutrons/cm^2/s$.

\begin{figure}[htbp]
    \centering
    \includegraphics[width = 0.8\textwidth]{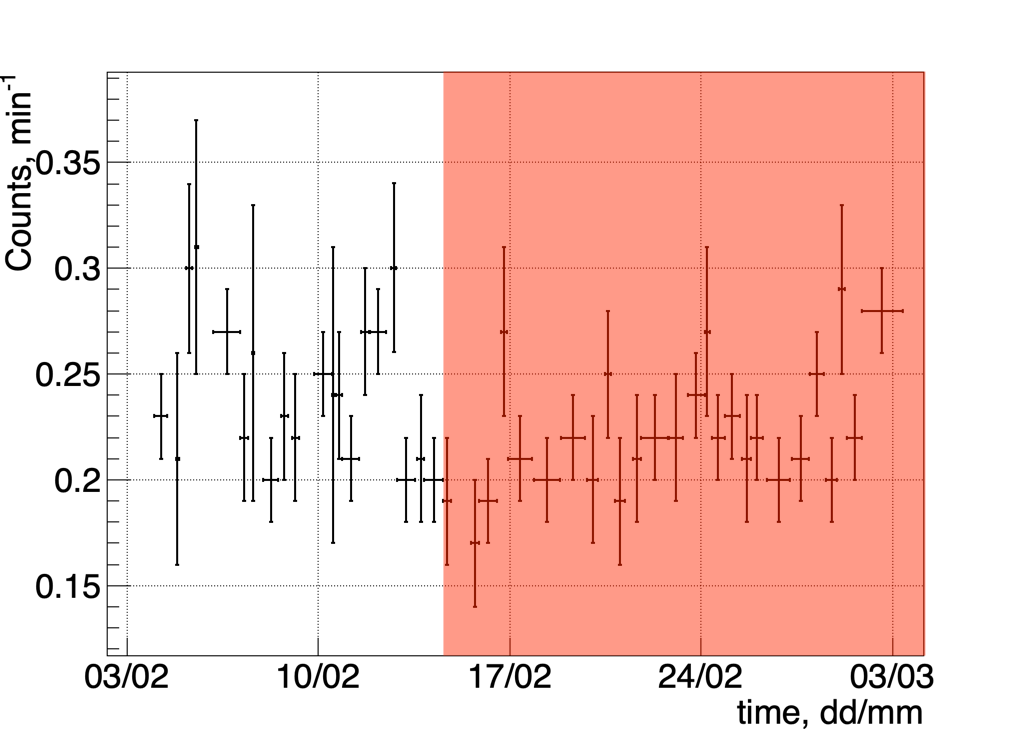}
    \caption{Count rate of neutron events obtained with the Bicron liquid scintillator detector; the red transparent rectangle represents the reactor ON period; horizontal error bars show the duration of each run; the last point on the graph corresponds to the long run after the water shielding of RED-100 was drained.}
    \label{fig:bicron_count_rate}
\end{figure}

Using the upper limit on the neutron flux and a neutron energy spectrum we estimated the associated background count rate in \cevns{} ROI of RED-100. Preliminary \gf{} simulation suggests that it is about one event per day, which is of the same order of magnitude as the expected \cevns{} signal. Therefore, possible variations of the neutron flux with the reactor operation have to be determined to estimate their potential influence on the RED-100 sensitivity to \cevns{}. For this purpose, the count rate of neutron interactions in the Bicron detector corrected per unit of live time was plotted versus the astronomical time (Figure~\ref{fig:bicron_count_rate}).

One may notice that the mean neutron count rate during the reactor OFF period is slightly higher than during the reactor was ON which is unexpected. The difference is $\sim10\%$  with about $2.3 \sigma$ significance. At the same time, it is unlikely that the reactor power ramping up can reveal itself as a reduction in the ambient neutron flux. We suppose that some technical preparations made by the KNPP staff a day or so before the reactor was turned on could affect that. For example, it might be due to increased boron concentration inside the primary circuit reactor water. There are several stations to measure its concentration around the RED-100 locations. Each station is equipped with a PuBe neutron source. Thus, increasing the boron concentration can reduce the part of ambient neutrons background associated with these sources. According to the WWER-1000 reactor operation technology, the boron concentration should be increased if the new fuel is installed which happened in our case. We also have a confirmation from the reactor authority that the boron concentration has increased in that period.

\section{Radon background}
\label{sec:radon}

The $^{222}$Rn level was controlled by two radon indicators RADEX MR-107 and MR-107+ with a nominal sensitivity threshold of 30 Bq/m$^3$. As it was reported in our previous paper~\cite{akimov2022red}, the measured radon level varied from the detection threshold of $\sim 30$~Bq to $\sim 100$~Bq with few peaks of up to $\sim 300$~Bq during the data taking.

Now, we have got the full dataset of the NaI[Tl] detector analyzed. The calculated count rate of NaI[Tl] in the full energy range can constrain variations of $^{222}$Rn level with time in both the water shield and the surrounding air. For that purpose \gf{} simulation was performed and $^{222}$Rn concentration was limited at 95\% confidence level based on the absence within the errors of the diurnal NaI[Tl] count rate variations. These concentrations are $189 \pm 17$ Bq/m$^3$ for the air and $3.2 \pm 0.3$ Bq/L for water if we don't take into account the solubility of radon in water. At the same time, the radon solubility in water is quite low (0.0093 mol/kg/bar~\cite{chase1998nist} at room temperature). The partitioning coefficient of $^{222}$Rn between pure solvent and air is about 0.24 at room temperature~\cite{jobbagy2017brief}. This can further reduce the possible variations of $^{222}$Rn in the water shield for an additional two orders of magnitude.

Having NaI[Tl] count rate behaviour which is more reliable than the domestic indicators data the latter were additionally constrained. Averaging the measured level of $^{222}$Rn in the air by RADEX detectors through the data taking period, we estimated its rate as $41 \pm 15$ Bq/m$^3$. Also, we didn't observe any correlation of the $^{222}$Rn count rate with the reactor operational status.

\section{Muon and muon-induced background}
\label{sec:muon}

Operating at surface RED-100 is exposed to the high flux of atmospheric muons~\cite{heusser1995low}. This background can be well identified and vetoed by the special blocking trigger described in our previous paper~\cite{akimov2022red}. At the same time, muons cause secondary backgrounds. First, they can generate secondary neutrons in the copper shield ~\cite{kneissl2019muon} and RED-100 has only a few centimeters of water protection from these neutrons. Second, and the most dangerous for \cevns{} study, muons generate a huge energy deposition inside LXe followed by a long and intense tail of delayed single electron (SE) signals background. Multiple coincidences of these SE events can mimic \cevns{} events~\cite{sorensen2018two, santos2011single, akimov2016observation}. It was shown previously~\cite{akimov2020first} that the latter kind of background dominated in the region of interest for the \cevns{} search.

\begin{figure}[htbp]
    \centering
    \includegraphics[width = 0.8\textwidth]{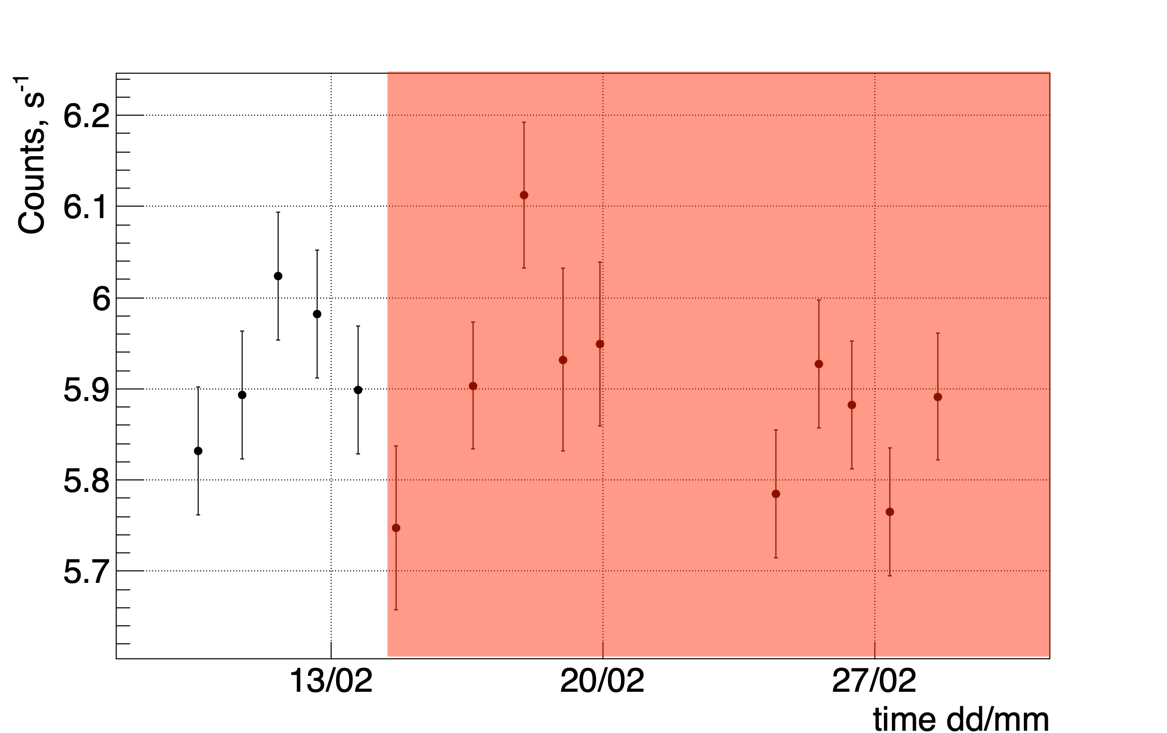}
    \caption{Muons count rate measured with RED-100 itself; red shadowed rectangle corresponds to the reactor ON period.}
    \label{fig:muons_rate}
\end{figure}

During the reactor OFF period and at the reactor turning ON the amount of water in the pools above the RED-100 location varied, which could cause changes in muon flux. Specially dedicated runs of RED-100 were carried out to measure muon flux within RED-100. The muon count rate observed in these runs is presented in figure~\ref{fig:muons_rate}, and it is about 7 times lower than in the laboratory test. We also didn't observe any correlations with the reactor operational status. Thus, we expect the muon-induced background to be independent of the reactor operational status.

\begin{figure}[htbp]
    \centering
    \includegraphics[width = 0.8\textwidth]{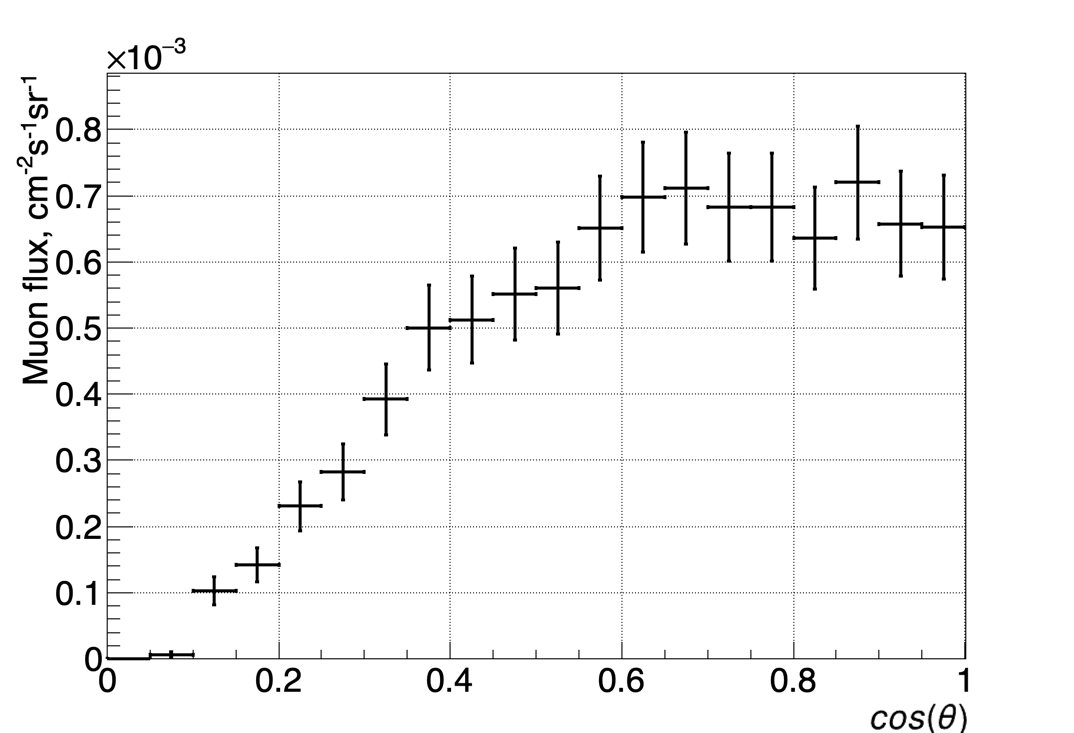}
    \caption{Muon angular distribution at the RED-100 location.}
    \label{fig:muons_angular}
\end{figure}

For the estimation of the muon-induced neutron background originating in the copper shield, the following procedure was performed. Muon tracks recorded with RED-100 were extracted by sampling 250~$\mu$s within the central part of waveforms (which have the full length of $\sim270~\mu$s) into 50 equal 5~$\mu$s pieces with the further reconstruction of their spatial coordinates. Points with energy deposition above a preset threshold were fitted by a straight line using the principal component analysis method. Then, for the tracks passing through the sphere which is concentric with RED-100 fiducial volume and has a radius of 13~cm (5~cm less than RED-100 internal radius) the polar and azimuth angular distributions were obtained. In figure~\ref{fig:muons_angular} the distribution of the cosine of the polar angle ($\theta$) is presented. Azimuthal distribution appeared to be isotropic in the RED-100 location. It should be noted that the RED-100 angular acceptance with this method is not uniform for all the polar angles. The efficiency of extracting horizontal tracks (90-degree polar angle) is zero since it is visible in RED-100 as a single but very intense point of interaction in the vertical direction. The efficiency becomes equal for the tracks with a polar angle below around 75 degrees ($cos(\theta) > 0.25$).

With this angular distribution, the number of muon tracks that pass through the copper shield but are invisible by the RED-100 detector was estimated with geantinos in \gf{} simulation. The average path in copper per such tracks was calculated. Then, using the neutron yield per cosmic muon of $2.1 \cdot 10^{-5}g^{-1}cm^2$ and mean energy of 8.9~MeV~\cite{kneissl2019muon} neutrons were randomly generated in the copper in \gf{} model. This resulted in about 30 events per day in \cevns{} ROI for RED-100. This background is expected to be stable following the muon flux stability.

\begin{figure}[htbp]
    \centering
    \includegraphics[width = 0.8\textwidth]{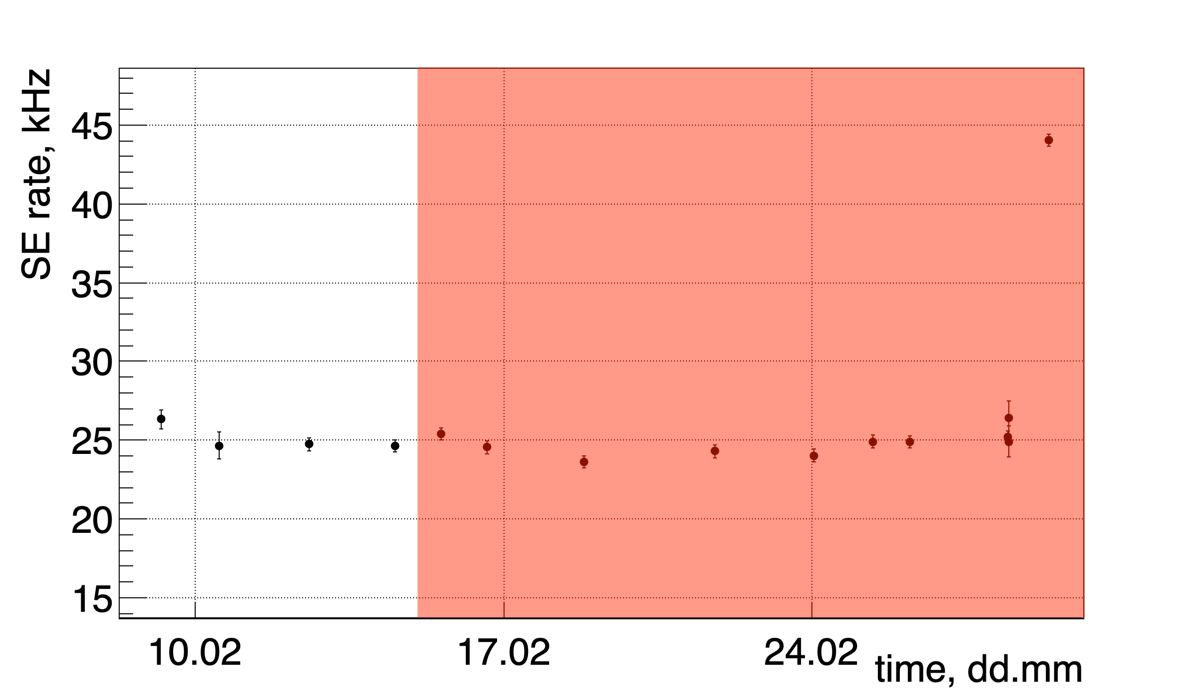}
    \caption{Single electron count rate measured with random trigger in RED-100; red shadowed rectangle corresponds to the reactor ON period.}
    \label{fig:se_count_rate}
\end{figure}

To check if the muon-induced SE background is stable during the RED-100 data taking period, the single electron rate was measured in special runs using a random trigger. This random trigger was generated with a 2 Hz frequency by the generator, independently from any events in RED-100 except for muons veto. Waveforms with a duration of 300 $\mu$s were recorded. Then, single electron clusters were found on those waveforms and their rate was calculated. This rate is presented in figure~\ref{fig:se_count_rate}. 

It was rather stable during the data taking period. The difference between the reactor OFF and ON periods is $1.9 \pm 1.2\%$. The count rate reduction was about a factor of 9 in comparison with our laboratory test. The rising of the count rate at the end of the data taking period corresponds to the water draining from the passive shield of the detector. The latter increased the amount of high-energy gammas passed through the detector which also contributed to the single electron noise. The higher suppression level with respect to the laboratory test of the SE background than that of the muon rate can be explained, particularly, by the passive shield around the detector at the KNPP which reduces the external background flux contributing to the SE noise rate.

\section{Conclusion}
\label{sec:conclusion}

In this paper, we presented the data analysis of the ambient background at the RED-100 location at KNPP. Having several supplementary detectors for the background study and monitoring we measured the fluxes of the most important backgrounds which will be used for the model of background in the \cevns{} search energy region of interest of RED-100. We have shown that the ambient background count rate didn't increase with the reactor turning ON for all measured types of background which could mimic \cevns{} rate.

At the same time, we observed that there was a decrease in the neutron background rate by about 10\% around the date when the reactor was on. Of course, it can't mimic the increase of the count rate with reactor ON since the downshifting in rate is observed. Furthermore, according to our preliminary computer modeling, ambient neutron flux provides about one event in the \cevns{} region of interest per day. Thus, $\sim10\%$ decreasing of its flux seems not important for the \cevns{} observation but has to be taken into account for the precise \cevns{} study.

We have also shown that the main source of background in the region of interest, a spontaneous single electron noise produced by large energy depositions in the detector, was stable and didn't depend on the reactor operational status. The count rate of this background decreased by a factor of $\sim$9 in comparison with our measurement in the laboratory conditions. This decrease was mainly due to the decrease of the muon background at the RED-100 location in comparison with that at the laboratory.

\acknowledgments

The authors express their gratitude to the State Atomic Energy Corporation Rosatom (ROSATOM) and the Rosenergoatom Joint-Stock Company for administrative support of the RED-100 project,
the JSC Science and Innovations (Scientific Division of the ROSATOM) for the financial support
under contract No.313/1679-D dated September 16, 2019, the Russian Science Foundation for the
financial support under contract No.22-12-00082 dated May 13, 2022, the administrations of the
National Research Nuclear University MEPhI (MEPhI Program Priority 2030), the National Research
Center “Kurchatov Institute”, the Institute of Nuclear Physics named after G.I. Budker SB RAS,
the Tomsk Polytechnic University (Development Program of Tomsk Polytechnic University No.
Priority-2030-NIP/EB-004-0000-2022) for support in the development of technology of two-phase
emission detectors. The work was supported by the Ministry of Science and Higher Education of the Russian Federation, Project “New Phenomena in Particle Physics and the Early Universe” FSWU-2023-0073. The authors are grateful to the staff of the Kalinin NPP for their comprehensive
assistance in conducting the RED-100 experiment, as well as the scientists from DANSS, $\nu$GeN, and iDREAM experiments at the Kalinin NPP, for assistance in organizing measurements.

\bibliographystyle{JHEP}
\bibliography{bibliography}

\end{document}